\documentclass{aastex}
\usepackage{spr-astr-addons}

\usepackage{graphicx}
\usepackage{txfonts}
\usepackage[]{hyperref}

\RequirePackage{color}

\begin{document}

\title{Solving the missing GRB neutrino and GRB-SN puzzles}

\shorttitle{Solving GRB-$\nu$ and GRB-SN puzzles}
\shortauthors{Fargion \& Oliva}

\author{Daniele Fargion}
\affil{Physics Department, Rome 1 University and INFN Rome1, Ple. A. Moro 2, 00185, Rome, Italy}
\and
\author{Pietro Oliva}
\affil{Electrical Engineering Department, Niccol\`o Cusano  University, Via Don Carlo Gnocchi 3, 00166 Rome, Italy}
\affil{MIFP, Mediterranean Institute of Fundamental Physics - Via Appia Nuova 31, 00040 Marino (Rome), Italy}

\begin{abstract}
 Every GRB model where the progenitor is assumed to be a highly relativistic hadronic jet whose electron-pair secondaries are feeding the jet's engine, necessarily (except for very fine-tuned cases) leads to a high average  neutrino over photon radiant exposure (radiance) ratio   well above unity, though the present observed average IceCube neutrino  radiance  is at most comparable to the gamma in the GRB one. Therefore no hadronic GRB, fireball or hadronic thin precessing jet, escaping exploding star in tunneled beam, can fit the actual observations.
A new model is shown here, based on a purely electronic progenitor jet, fed by  neutrons  stripped from a neutron star (NS)  by tidal forces of a black hole or NS companion, it may overcome these limitations. Such thin precessing spinning jets explain unsolved puzzles such as the existence of the X-ray precursor in many  GRBs.  The present pure electron jet model, disentangling gamma and (absent) neutrinos, explains naturally why there is no gamma GRB correlates with any simultaneous TeV IceCube astrophysical neutrinos. A thin persistent electronic beaming, born in an empty compact binary system has the ability to offer the answer for a sudden   engine (the thin jet) whose output may be comparable, off axis, to $10^{44}$--$10^{47}$ erg s$^{-1}$. The jet power is fed by a stripped neutron mass skin by tidal forces. The consequent jet blazing to us on axis occurs within the inner jet cone beammed by a spiral charged ring at highest apparent output.
  In rare cases, the NS, while being stripped by the BH companion, will suddenly become unstable and it will explode and shine during the GRB afterglow, with an (apparent) late SN-like event birth. Primitive SN outer chemical mass shells, should be retro illuminated by such a NS  explosion, re-brightening the relic nuclei as in a SN-like spectral line signature. To disentangle SN from NS explosion we note that only radiative shining due to Cobalt and Nichel decay,   present in most SN, will be absent in present NS explosion. Recent IceCube-160731A $\nu_\mu$ event with absent X-$\gamma$ traces confirm the present model.
   \end{abstract}


\maketitle

\section{Introduction} \label{sec:intro}

Gamma ray burst (GRB) physics represents today a half-century (1967--2016) unsolved puzzle which brings together a long list of unanswered questions related to the many faces a GRB can show.
 The main popular fireball model and its modern variations are always doomed to fail in front of a key lethal unanswered question: how we do explain the existence of tiny X-ray precursors (present in hundreds of GRBs) seconds or minutes before the huge apparent gamma explosion? No fireball nor any one-shot fountain model even try to face this reality or seems to be comfortable with the existence of precursors. Maybe the time has come to embrace a change.

One of the most important puzzles to recall is: how is it possible that a huge GRB (apparently isotropic) power $P_{\rm GRB}\sim10^{53}$~erg~s$^{-1}$ can sometimes coexist (see i.e., \cite{1998Natur.395..672I, 2014A&A...567A..29M}) with a late correlated supernova (SN) event of the typical order of $P_{\rm SN}\sim10^{44}$~erg~s$^{-1}$, a power billion times weaker?
Indeed, this question represents only the tail of a long chain of mysteries surrounding the nature of GRBs. First of all, because of the fast millisecond-second scale of GRB variability, how could any corresponding compact source emit at MeV energies any apparent spherical GRB luminosity $P_{\rm GRB}\gtrsim10^{51}\div10^{53}$~erg~s$^{-1}$ several orders of magnitude above Eddington limit for such objects $(\sim10^{38}$~erg~s$^{-1})$? In such a model photon scattering will lead to the birth of electron pairs so dense and opaque that they will definitively  screen off and shield the GRB self prompt compact spherical explosion. Moreover no GRB show just a single bang (as in a SN), on the contrary the most of them show a sequence of peaks in gamma.
  \begin{figure*}
 \centering
  \includegraphics[scale=0.95]{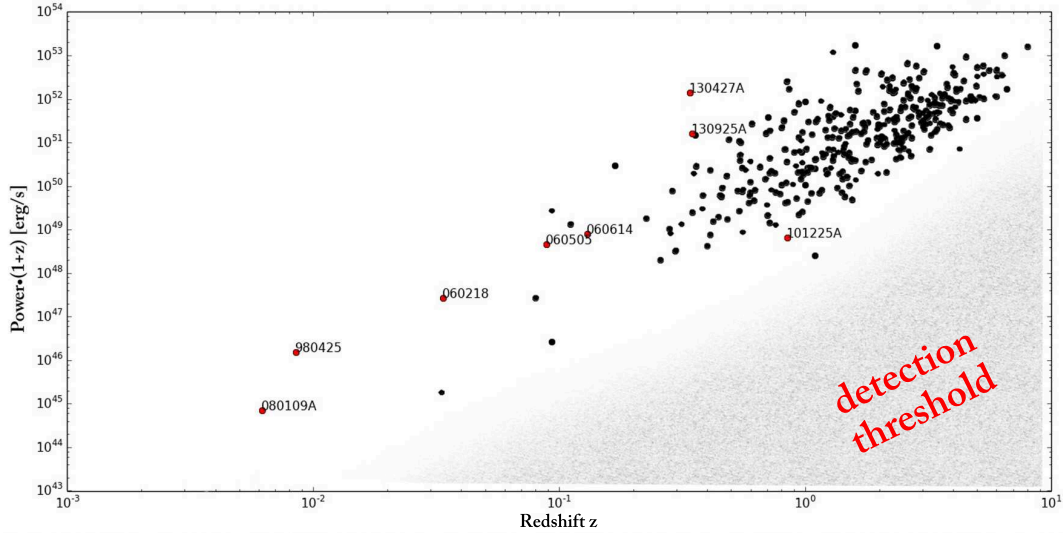}
\caption{The complete sample of GRBs with known redshift plotted against their relativistic invariant peak power (evaluated in a standard expanding cosmic model, assuming isotropic radiation) shows many orders of magnitude increment with its redshift. The rarest soft GRBs, as the nearest ones, have to be very abundant also at far redshift, but they are hidden by their weak detection threshold; the far away GRB are located in the largest volumes and in richest sample, where the most rarely aligned $\gamma$ jet might be pointing to us emerging as the brightest and hardest (and  often mostly variable) ones; their thinner jet beam whose harder core is narrow because the most energetic UHE electrons showering in gamma  shine with brightest luminosity while the wider cone that are fed by lower energetic electron pairs may naturally explain the longer life X afterglows and the apparent anti-copernican evolution around us. Also the hard-luminosity connection found in Amati diagrams has a natural explanation in the beamed relativistic jet cones structure. Let's also remind, among the puzzles, that UHECR distribution still appears isotropic and uncorrelated with sources, even considering magnetically-induced alignments (see i.e. [\cite{2012APh....35..354P}]).}
\end{figure*}\label{fig:06}

The early (1980--2000) ``fireball'' model [\cite{Cavallo01071978, 1986ApJ...308L..47G, 1986ApJ...308L..43P, 1992MNRAS.258P..41R, 1994ApJ...430L..93R, 1993ApJ...418L...5P, 1997ApJ...489L..33W, 1997ApJ...489L..37S, 1997ApJ...478L...9V, 1999ApJ...517L.113C}] tried to explain that the  sea of electron pairs from a GRB will spread out and dilute in a sphere, the so-called fireball, hence cooling the photons in an adiabatic expansion  from MeV to keV energies. The model then foresaw that when the pair-sea shell would have become sufficiently diluted and transparent, these keV photons (ejected and scattered by these ultrarelativistic electron pairs) would reach us boosted at MeV energies like the ones observed in GRBs.
Since the Beppo-SAX identification and discovery of the high cosmic redshift of some GRBs with extremely high luminosity \cite{1997ESASP.382..179P, 1997SPIE.3114..186F, 1998A&A...332L..29F} this simple isotropic model depicting ``spherical'' GRBs failed, mostly because of the observed highest GRB integrated energy $(E_{\rm GRB}\gtrsim10^{54}\,\mathrm{erg})$ which is comparable or larger than the same source budget allowable energy  mass, a mass derived and constrained by the object's Schwarzschild radius (fixed or constrained by its variability). Clearly, such an energy budget paradox could not be solved by an increase of the GRB mass and its Schwarzschild radius because of the subsequent increase of the variability time scale in disagreement with  the observed fast ms GRB timescales.

Subsequently in 2000, most authors abandoned the spherical fireball model and turned to a mildly beamed jet-explosive fountain model with a $\Delta\Omega/\Omega\sim10^{-3}$ ratio \cite{1999ApJ...517L.109S, 2000ApJ...529..146E, 2000AIPC..526..514M} while the inner (random) variability (peaks and sudden re-brightening)  of the GRB luminosity was explained assuming that  the fountain jet would hit  relic shells of matter around (but external) the GRB, where shock waves revived the GRB luminosity.
Unluckily for fireball believers, this ad hoc model was and still is not able to explain the multi-peak structure of some GRBs: to face this variability and to  keep alive the fireball model several authors considered the far external relic shells of the exploding GRB  star as the additional onion-like screens where, by scattering of the expanding shock waves, the explosive luminosity re-brightens several  times. Obviously this process, fireball defenders said, must open the fireball fountain jet into an increasingly  spread out spherical explosion with a more and more diluted luminosity. Several GRBs on the contrary proved an opposite growing peak luminosity trace. Moreover each onion shell in such models must be  diluted enough to transfer outside the GRB shock wave but not too diluted for being transparent to the scattering: such a fine-tuned GRB dressing for fireball is purely ad hoc and unexplained. In particular the fireball one shot model is totally incapable of describing and justifying the early  X-ray precursor \cite{Fargion:2001xf, 2001foap.conf..347F} present in a significant fraction $(\sim 7\%\div15\%)$ of GRB curves up to date. These earliest bright X-ray flares may hold a million times the SN luminosity even several minutes (ten minutes for GRB 06124) before the main (billion times brighter) harder GRB event. Moreover, the wide beaming of the fountain $(\Delta\theta\sim10^\circ\div15^\circ)$ is assumed ad hoc and the single-shot model cannot describe some observed long life and ``day after'' re-brightening GRBs, nor the several week X-ray afterglows. Moreover the fireball model is unable to justify the apparent ``conspiracy'' that makes GRB more and more (in apparent) brightest power at larger and larger redshift, in a spread of apparent luminosity of nearly a factor a billion discussed below: a beaming factor of just a thousand as in fireball model, cannot explain more than a thousand in luminosity range variability. On the contrary a thinner precessing jet  whose solid angle is a million or billion times smaller, may embrace a million or a billion luminosity variability. The same plot play a role in making (apparent) harder and harder the GRB spectra with the more and more distance (and red-shift). Naturally, we are observing a statistical geometry evolution that allows the most distant and richest sample to have the most aligned and thinner jets pointing towards us, while the nearer (smaller cosmic volume) and rarer GRB are usually off-axis and they shine with low fluxes.

\section{An anti-Copernican GRB Luminosity evolution?}

Among the contradictions of all GRB one-shot models stand the apparent conspiracy of GRB luminosity
around us: nearby (lowest red-shift) GRBs show on average a peak luminosity and a soft energy spectra versus the much brighter
and harder luminosity of far away (large redshift) GRB events.
The conjure or the apparent luminosity evolution  is so fast that it suggest that we (in our local Universe) are at the
center of the Universe. There is not any ad hoc  luminosity evolution that may explain such a
  sudden ($z\gtrsim0.01$) growth in spectra and luminosity evolution.
  This result is manifest in Fig.~\ref{fig:06}, and it calls for an explanation.
A wide fountain and a marginal beaming as in a fireball model cannot explain such a factor of a billion in luminosity spread;
a very thin beaming (as will be discussed below within  a millionth or less of steradian solid angle) spinning and precessing jet
 has a characteristic  angle  linked to the peak electron  Lorentz factor above thousands value:
the inverse of the solid angle and the apparent luminosity grows as large as the square of the Lorentz factor (of highest energetic electrons).
Of course also a hierarchic cannibal event between binary compact objects may play a role, showing new rare powerful jet with wider distances and volumes.
However as it is well known binary  (Schwarzschild or Kerr neutral) BH merging systems are ejecting only gravitational waves (GW).
Therefore only (or mainly) neutron star merging, as discussed below, in BH-NS or in NS-NS systems are a guaranteed source of electromagnetic radiation and the NS are a well-bounded amount of mass-energy. Therefore even if the
GRB event is fed by NS-NS or NS-BH  binary merging, even for large and large BH, the outgoing energy budget in GRB
is nearly fixed and bounded by the NS mass. The huge luminosity variability is due to the very thin beaming geometry
associated to tens-hundred GeV electron pair jets, not to any hierarchic growth of objects.

 \begin{figure}[h]
 \centering
 \includegraphics[scale=.39]{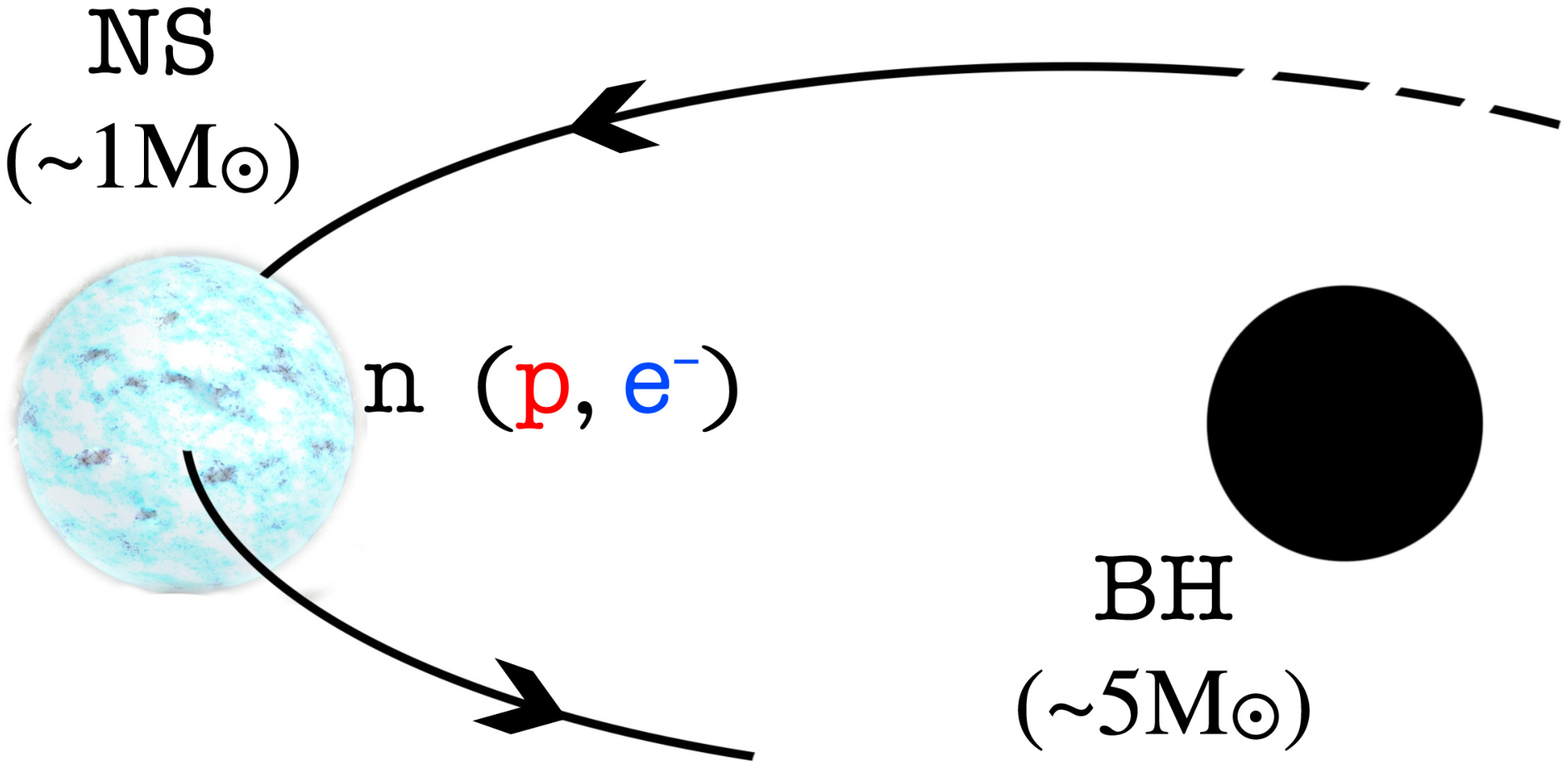}
  \includegraphics[scale=.33]{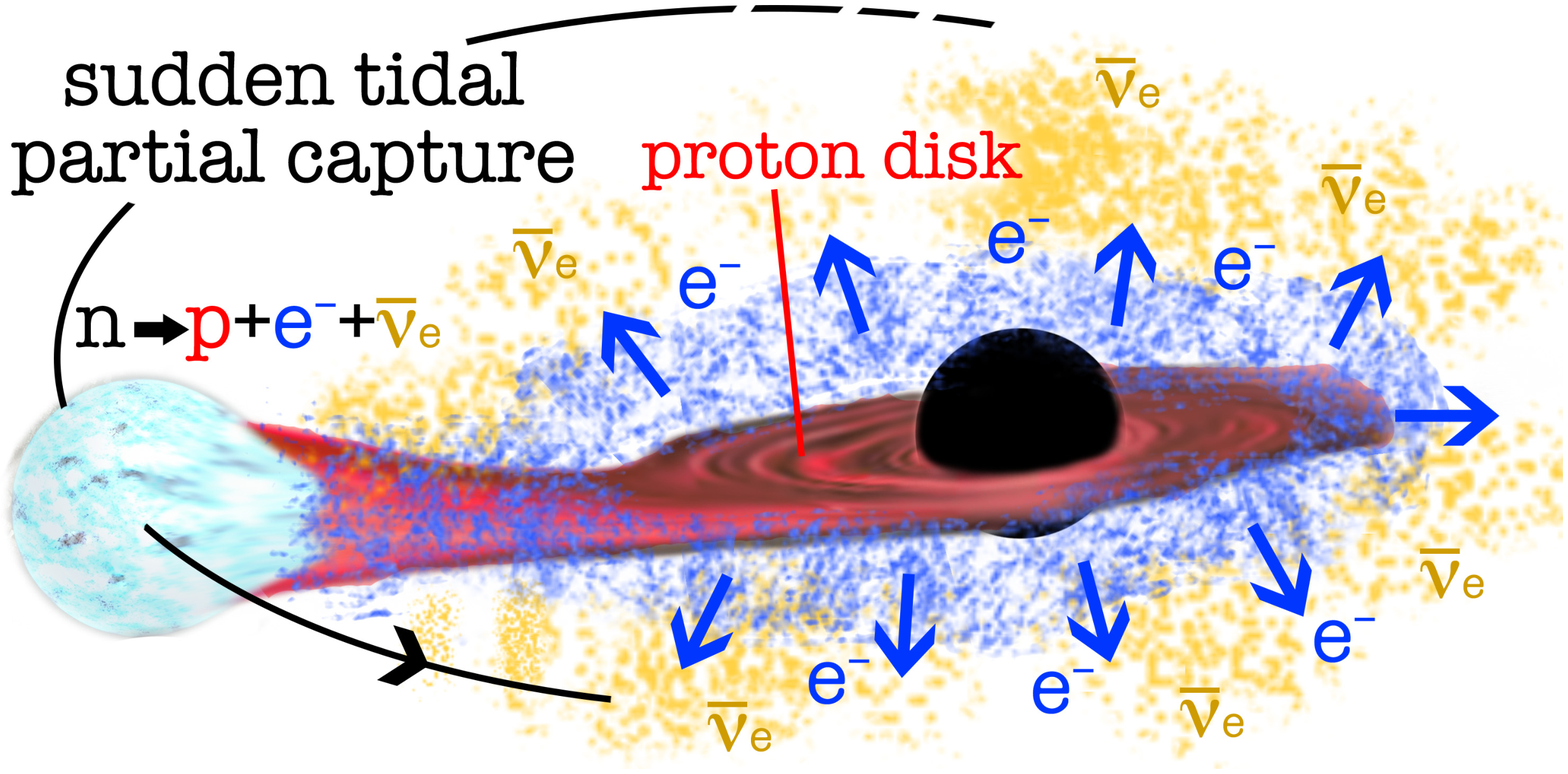}
\caption{\textit{top}: Neutron star (NS) orbiting in an elliptical eccentric trajectory, skimming a black hole (BH) companion object;\newline
\textit{bottom}: NS suffering a tidal force able to strip neutron dense matter along an accretion disk. The neutron in free fall start to decay leading to a nearly (unmoved) proton tails, a free spherical evaporating $\sim$ MeV beta decay $\bar{\nu}_{e}$ and an almost similar cloud of $\sim$ MeV electrons.}
\end{figure}\label{fig:01}

\section{Precessing and spinning of thin decaying $\gamma$ jet} \label{sec:prec_jet}

In order to overcome the GRBs puzzles we proposed since 1994 \cite{1994dsu..conf...88F, Fargion1995269} a model to describe both GRBs (and/or SGRs) based on the blazing of a very thin $\gamma$ beamed jet $(\Delta\theta\sim0.1^\circ\div0.02^\circ)$, $\Delta\Omega/\Omega\lesssim 10^{-6}\div10^{-8}$ whose birth was associated to tens GeV electron pairs showering via inverse Compton scattering (ICE) into MeV-GeV photons \cite{{1997ZPhyC..74..571F}, 1998PhyU...41..823F}. Our precessing-spinning $\gamma$ jet was assumed fed at a low power (fitting today SGR or AX-PSRs) in our galaxy $(P_{\rm SGR}\sim10^{38}\,\mathrm{erg}\,\mathrm{s}^{-1})$ or, since 1998 \cite{1999A&AS..138..507F}, also at highest power as large as a SN powering and beamed jet for cosmic GRBs $(P_{\rm GRB}\sim\gamma_e^2\mathrm{P}_{\mathrm{SN}}\simeq10^{50}\div10^{54}\,\mathrm{erg}\,\mathrm{s}^{-1})$.

 \begin{figure}[t]
 \centering
   \includegraphics[scale=.38]{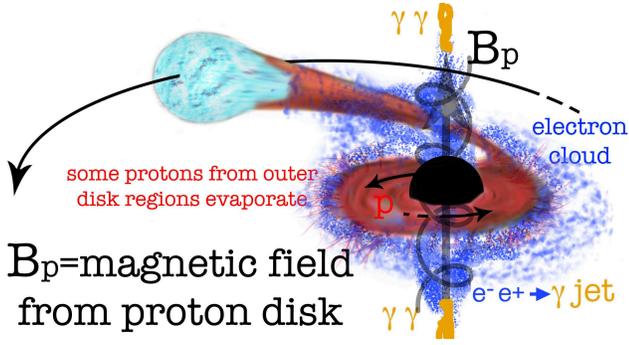}
\caption{Protons follow their ring trajectory while in $\beta$-decay forming a net charged current and a huge aligned magnetic field $\mathbf{B}_{p}$. The evaporating electrons are easily captured and aligned along $\mathbf{B}_{p}$; their crowding at the North and the South Poles create a huge electrostatic gradient that makes a powerful linear active accelerator: an electronic jet arises and ejects electrons and/or electron pairs by bremsstrahlung as well as photons (by inverse Compton scattering and synchrotron radiation);  the thin spinning and (by tidal gravity forces) precessing jet, drives a collinear $\gamma$ jet making a blazing dance by its geometry beaming \cite{2006ChJAS...6a.342F, Fargion:2006we}. Once on axis, we are dazzled and we call it a GRB event.}
\end{figure}\label{fig:02}
  Late GRB jet power, decaying with a power law $\approx t^{-1}$, may shine as an nearby exhausted soft gamma repeater (SGR) jet source where the output power is correlated with a thousand year time delay with the early GRB and present SGR output.
   The geometrical spinning and precessing of the thin GRB-SGR jet naturally explain the huge GRB variability and the quasi-periodic behaviors found in well recorded SGR events.
  In the present model discussed below, the feeding of  stripped matter of a NS by a black hole (BH) or a NS companion, is shining energy:
   indeed stripped neutrons and protons condense into a charged spiral ring that is paying the energetic output budget
   to eject a thin collimated, spinning and precessing electron jet, at $10^{44}-10^{47}$~erg~s$^{-1}$ output; moreover the bending geometry of the electron jet (by bending magnetic fields of the accreting ring and the BH spin) and its consequent beamed variability, explain the huge and fast GRB-SGR luminosity.
The fact that the neutron by NS star stripped matter and its decayed  protons will follow the spiral geodesic around the BH or NS cannibal companion  (while the electrons and neutrinos will not) will lead to a charged ring and a sudden collimating magnetic field.
This decaying neutron-$e^+e^-$ pairs-proton ring, which is also pulsating, can shrink the magnetic lines and it can force  the electrons trapped in the poles into an ultra-relativistic jet which will later create the observable gamma jet.
This novel electronic model  is able to avoid the pion progenitor and the overcrowded neutrino tails foreseen in all hadronic GRB models explaining GRB-$\nu$ absence.

\subsection{GRB with SN event}

 In some occasion  such an electronic jet model formed around the BH, or heaviest NS companion, may also lead to an explosion of the relic stripped NS binary, which is now unstable because of the spoiled and stolen external  weights.

Because of the extremely beamed angle $(\Delta\Omega/\Omega\sim10^{-6}\div10^{-8})$ these apparent luminosity, if seen in-axis  by the observer, would shine apparently as bright as a $\tilde{P}_{\rm SGR}\sim10^{44}\div10^{46}\,\mathrm{erg}\;\mathrm{s}^{-1}$ while $\tilde{P}_{\rm GRB}\sim10^{50}\div10^{54}\,\mathrm{erg}\;\mathrm{s}^{-1}$.
The lifetime of the jet has been assumed not to be a one-shot event (as the fireball model does).
On the contrary our thin precessing and spinning jet has a characteristic decay life about $t^{\rm GRB}_{\mathrm{decay}}\simeq\left(t/t_{0}\right)^{-1}$, where $t_{0}\simeq 3\times10^4$ s. This half-a-day timescale was chosen to connect, by a time decay law $P\sim \left(t/t_{0}\right)^{-1}$ the highest GRB output to late, thousand years later, less powerful relic, almost steady (Galactic as SS433) Soft Gamma Repeaters, SGRs.

Despite being able to explain even the X-ray precursor (by a peripherals skimming shine of the jet to the Earth, before the main jet blazes as a GRB) and the late GRB re-brightening through simple geometry beaming, the precessing jet model unifying GRB and SGRs was (and it is) often  underestimated or un-noticed \cite{1999A&AS..138..507F}  since twenty years.

\subsection{Hadronic jet feeding a fireball lepton--$\gamma$ jet} \label{sec:fireball}

The fountain-fireball model was --- and is --- based on shock interacting shells of hadrons (UHECR at PeVs$\div$EeV, protons and nuclei) leading to neutral pions $(\pi^0\to2\gamma)$ as well as to charged ones $(\pi^\pm)$ whose final decay results in electron pairs, the ones that later will shine in $\gamma$ in the GRB and a rich tail of neutrinos $(\nu_e, \nu_\mu, \bar{\nu}_e, \bar{\nu}_\mu)$ as well. There is also the possibility to feed pions by UHE nucleons and nuclei interacting with photons in flight.   Also more violent charmed hadronic reactions lead to prompt secondaries as the ones above. In this context the most popular fireball model foresees a comparable trace of $\gamma$  luminosity under the form of GRBs with respect to a neutrino radiance, as they were just secondaries of charged pions in decay in vacuum space.
Nevertheless, we repeat, GRB occur in dense stella shells in fireball model.
Naturally, because of the photon-photon interaction and/or IR-tens TeV opacity most of highest TeVs photons degrade and decay into MeV$\div$GeV ones (directly at their source or along their cosmic flight). This is not the case for tens TeVs or PeVs complementary neutrinos that may reach us unabsorbed showing (in this popular and ideal fountain-fireball model) the same radiance imprint of the partially absorbed  gamma observed in GRBs. As we shall comment, the transparent pion decay in flight, in fireballs, is a wishful  chain of events, mostly very unrealistic because most of the onion shell barrier encountered by the fireball jet will be (mainly at the inner core) opaque to photons but not to neutrinos. Photons will fed the kinetic energy of the barrier shells while UHE neutrinos will  escape with nearly no  losses. If the inner star core shells are opaque even to the neutrinos then only the rare interacting UHE neutrinos, making UHE penetrating muons at the external edges, may feed the GRB with electromagnetic secondaries, while most of the primary neutrinos will export to us  much more energy than gamma in GRB anyway. In conclusion the ratio gamma - neutrinos comparable to the unity is  a ``chimera".

The so-called Waxman-Bahcall (WB) limit or bound \cite{1999PhRvD..59b3002W}, which connects ten EeV cosmic ray (CR) radiance ($\Phi_{\rm CR}\sim$10~eV~cm$^{-2}$~s$^{-1}$~sr$^{-1}$) with average cosmic GRBs one $(\Phi_{\rm GRB}\sim\Phi_{\rm CR})$, constrains  the expected cosmic tens TeV$\div$PeV GRB neutrinos (GRB$\nu$s) at similar GRB energy radiance.
Indeed, the expected WB neutrino signal didn't arise with any correlated GRB yet, or it might be rarely $(\sim1\%)$ arose as a possible precursor. The absence of any prompt GRB--$\nu$ correlation represents a remarkable failure of any one-shot fireball version, even the most beamed one. No room for one-shoot GRB neutrino and gamma event \cite{2014arXiv1408.0227F}. Furthermore, any hypothetical dark or hidden population of GRB should not be considered, for this would call for a higher and higher ratio $(\Phi^\nu_{\rm  GRB}/\Phi^\gamma_{\rm  GRB}\gg1)$ while the observations are telling us $(\Phi^\nu_{\rm GRB}/\Phi^\gamma_{\rm  GRB}\sim1)$ \cite{2012Natur.484..351A, 2016arXiv160106484I}.

 \begin{figure}[t]
 \centering
 \includegraphics[scale=.32]{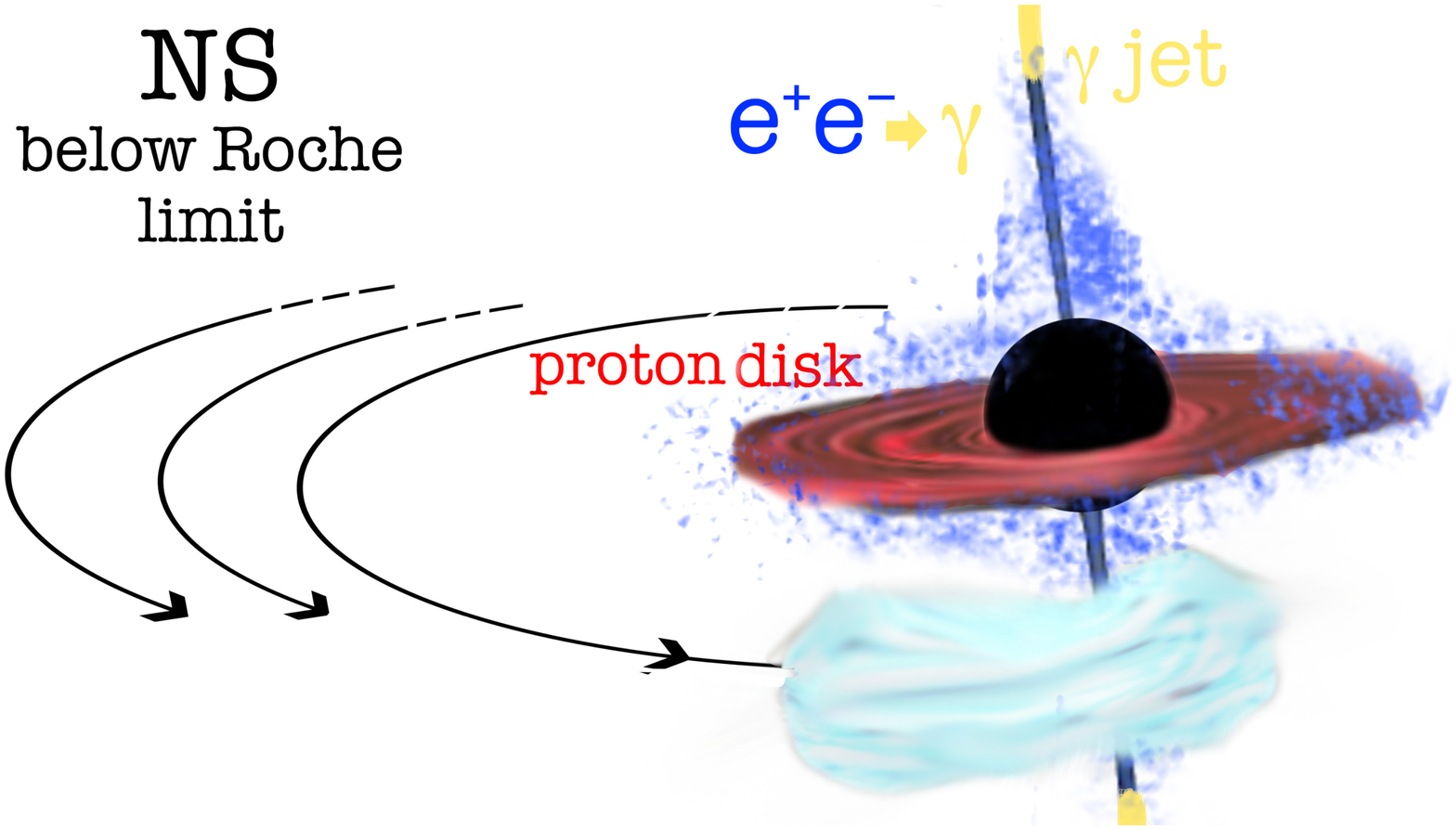}
 \includegraphics[scale=.27]{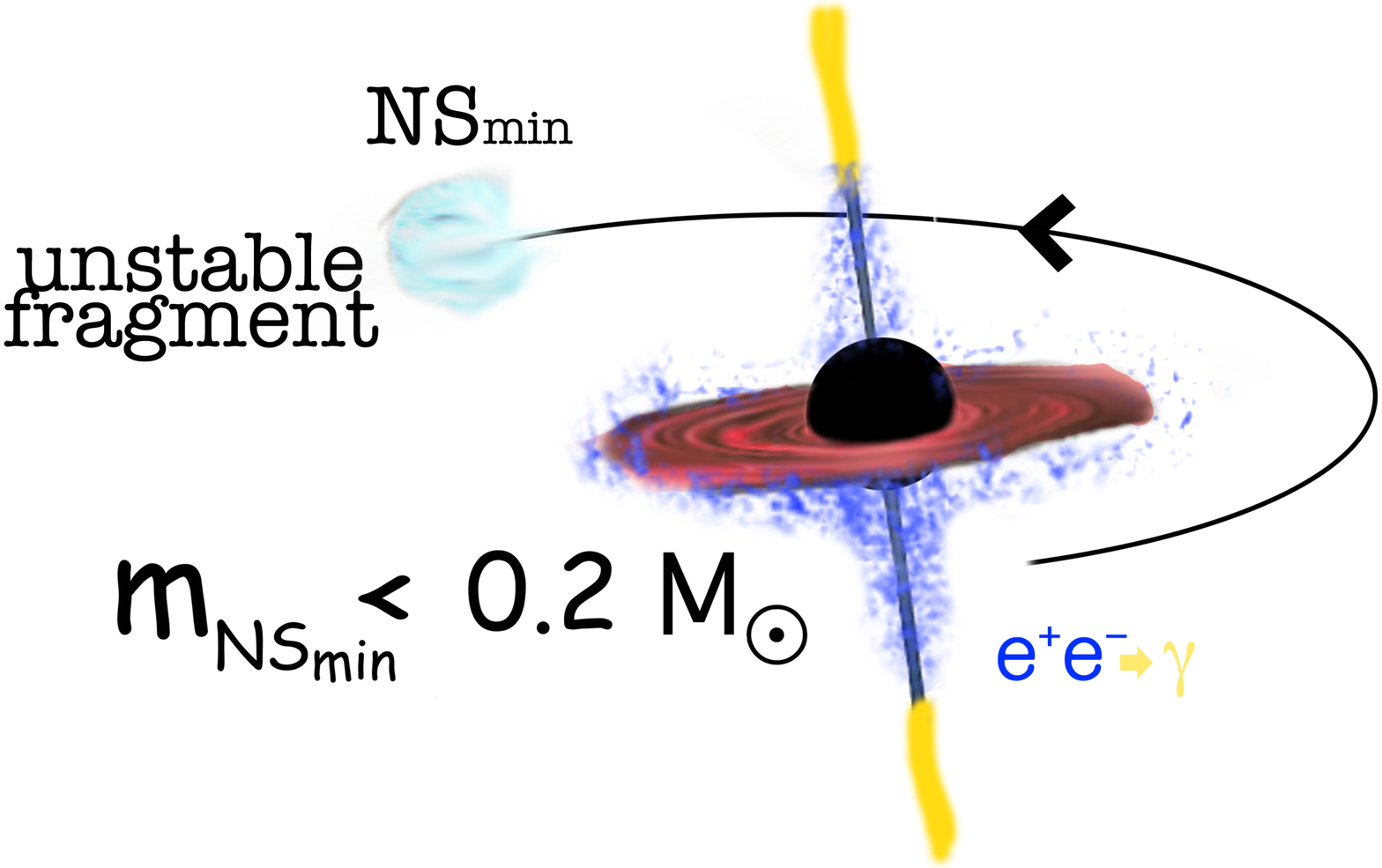}
\caption{\textit{top}: while in spiral trajectory the NS is sometimes too much bent and tidally disturbed by the BH up to lose an important fraction of its mass in the ring. It may also be a more quite serene and steady NS strip to lighter and lighter relic mass (it may be also that the final NS is eaten in a prompt step by the BH);\newline\textit{bottom}: anyway the survived NS fragment may become unstable (mostly below a minimal NS mass $m_{\mathrm{NS}_{min}}\lesssim0.2\,\mathrm{M}_\odot$).}
\end{figure}\label{fig:03}

In our thinner precessing jet we might solve the huge apparent GRB power spread puzzle in a first approach because of the ultra-relativistic beaming and the consequent thin beaming angle: the higher the energy, the thinner the jet cone and thus the rarer the blazing, which of course explains why we have observed (at tens to hundred keV) thousands of  GRBs, a few hundred GRBs at a MeV to tens MeV, a few dozen at a hundred MeV to GeV energies and only few rare events at a hundred GeV, the beming explaining their rarety. The precessing jet model can also shine in an almost cyclic fashion (like SGRs) and might blaze partially as a rare precursor, ruling out the mysterious 10\%$\div$20\% GRB events with precursors.
In principle a  thin relativistic beaming may explain that TeV neutrinos are so beamed that their shining inside the wider X-$\gamma$ cones happens very rarely. Furthermore, this requires a prompt $\nu$ detection with a fast follow up in X-$\gamma$ range. The first attempts (see next section), have failed.

\subsection{An hadronic or electronic precessing jet?}

We admit that our precessing $\gamma$ jet was originally based on hadronic-UHECR primaries, leading to PeVs $\mu^+\mu^-$ secondaries whose decay in flight were able to escape and survive the eventual opaque stellar mass layer and photosphere of a SN explosion \cite{2005NCimC..28..809F}. In addition, the same $\mu^+\mu^-$ shined in $\nu$, $\bar{\nu}$ at higher and higher than unity  ratio respect to photons; this applies for the following reasons: if GRB's $\gamma$ are made by relativistic electrons radiation and if the GRB jet are originated by UHECR hadrons inside the collapsing star, than only a small fraction of the UHECR energy radiance is able to escape the matter barrier in the form of secondary final $\gamma$ constituting the GRB. Most of the hadron jet energy is dispersed and wasted inside the baryonic shell kinetic energy and its temperature  along  the jet shock wave propagation.
  The basic  huge absorption of any electromagnetic traces respect to neutrino ones is a severe argument against any hadronic GRB origination. Present low (or missing) neutrino records in IceCube with respect to same observed gamma radiance in nearly a thousand GRB probe it \cite{2016ApJ...824..115A}.
 \begin{figure}[t]
 \centering
   \includegraphics[scale=.3]{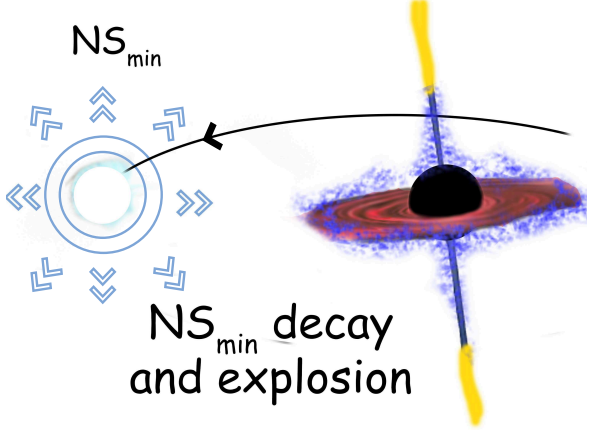}
\caption{Unstable NS suddenly evaporate its surface by free neutron $\beta$-decay toward a catastrophic NS explosion similar or even more energetic that a SN one.}
\end{figure}\label{fig:03b}

\subsection{The absence of $\gamma$-X signal from IceCube-160731A}\label{sec:lack}
The very recent prompt search of an electromagnetic trace by an astrophysical candidate event IceCube-160731A have proven the embarrassing absence of any optical (H.E.S.S.), X-$\gamma$ (Swift, Fermi, Agile) correlated signal (see \cite{report}). A first and rough estimate of downgoing energetic $\nu_\mu$ neutrino at $\sim100$ TeV in a km$^3$ IceCube has a pure probability to interact ($\sim10^{-4}$). therefore, the consequent expected IceCube-160731A released energy  is $\sim10^{18}$ eV over a km$^2$ area, or $10^8$ eV/cm$^2$ energy fluence. On the contrary the electromagnetic bounds in Swift (as well as in Fermi and other detectors) are as low as 1 eV/cm$^2$ energy fluence. The main consequence is that even no a part of a million is related in $\gamma$ respect to $\nu$ signal. Therefore, there are no hadronic jets in GRBs (or worse no clear understanding of IceCube astropysical $\nu$ nature, \cite{2015arXiv151208794F}).

 \subsection{Cosmic rays and hadronic jet surviving analogy}\label{sec:ana}

 To depict the analogy in a more clear way let's recall the CR metamorphosis along their flight inside the Earth atmosphere, which is a ten meters water equivalent (w.e.) screen: at ground level only a small amount of the CR energy is observable under the form of electromagnetic secondaries ($e^\pm$, $\gamma$).
Most of the surviving electromagnetic traces are indeed $\mu^+\mu^-$, whose energy radiance is already suppressed by two orders of magnitude with respect to the primary GeV $p$ (nuclei) at the top of the atmosphere.  Most of the relic energy is lost as heat and as kinetic energy spread by CR showering in air. A large fraction of the surviving CR trace is represented by the atmospheric neutrinos at a hundred MeV that exceed by 3$\div$4 orders of magnitude the corresponding MeV $\gamma$ component arriving at sea level, although in very special fine-tuned cases of EeV airshowers we can find a great electromagnetic component comparable to the $\nu$ one on the ground. In general the surviving atmospheric neutrino secondary tail exceeds by many orders of magnitude the corresponding electromagnetic component (mainly muons) while crossing the hadron barrier along the jet propagation.

To be more quantitative let's recall the ratio between $\nu$ and the electromagnetic tail of atmospheric CR both on the ground and in deep kilometer-underground detectors as well as across the Earth (for neutrino event rates in different scenarios see i.e. \cite{2012ApJ...758....3F}).
Atmospheric muons or $e^\pm$, $\mu^\pm$ from $\nu_{\mu, e}$, $\bar{\nu}_{\mu, e}$ are the observable electromagnetic traces in the last case: $\Phi_{\rm CR}/\Phi_\nu\simeq\Phi_{\rm CR}/\Phi_{\mu^+\mu^-}\gtrsim10^2$ on the ground; $\Phi_{\rm CR}/\Phi_{\mu^+\mu^-}\gtrsim10^8$ in underground detectors; $\Phi_{\rm CR}/\Phi_{\mu^+\mu^-}\gtrsim10^{14}$ in case of up-going signals \cite{AAFargion02, 2004ApJ...613.1285F}. The corresponding shields are namely 10 m w.e., 2 km w.e. and $10^5$ km w.e. In general the ratio between $\Phi_{\rm CR}/\Phi_\gamma$ is related to the ratio between the baryon barrier size $D_b$, the propagating lepton $\mu^+\mu^-$ distance $l_\mu$ and the interacting and propagating $\nu_\mu$, $\bar{\nu}_\mu\to\mu^+$, $\mu^-$.
In summary, the ratio $\Phi_{\rm CR}/\Phi_{\mu^+\mu^-}$ is related to the surviving muons and the propagating distance:
$\Phi_{\rm CR}/\Phi_{\mu^+\mu^-}\simeq e^{-D_b/l_\mu}$ and for largest baryon barrier ($D_b\gg12$~km) the muons arise by the appearance of high energy atmospheric neutrinos interacting with matter.

The lowest ratio (in first approximation) between a survived neutrino over a gamma average  GRB radiance (assuming a dozen km size rock shell along the hadronic jet trajectory) maybe estimated assuming (as for IceCube) a primary prompt 30 TeV neutrinos whose most penetrating secondaries (the muons) escape as well after tens km rock they are shining outside the shell as muon first and later on as electron pairs and gamma: $\Phi_{\nu}/\Phi_{\mu^+\mu^-}\simeq l_{\nu}/l_\mu$, above ten thousand. In conclusion the minimal ratio of neutrino over gamma radiance should be around ten  thousand and not one, if GRB are hadronic in primary nature.

\subsection{Where is the gamma radiance lost?}

If in the hadronic GRB jet, a large fraction of the gamma output is lost in opaque shells, one may wonder
that this is impossible because the energy conservation is lost. Indeed in the sun the radiation is both in photons and in neutrinos. Why should it not be the same in GRBs?
The reason is that the solar photons are in late thermal equilibrium stage while GRB photons are out of equilibrium.
Therefore where did the gamma energy fade (with respect to the neutrino one)? We believe that in
any hadronic GRB (outside of a fine-tuned case where the external shell are just transparent ad hoc)
 a large part of the gamma energy should be absorbed by the baryon matter while being scattered
and-or being absorbed, accelerating the shell masses in the form of kinetic shells.
The explosive kinetic shell masses as well as part of the survived cosmic rays (escaped along the jet) might contain the primary
hadron and gamma energy, while the neutrino component (born at the same inner sources) will suffer a negligible depletion, surviving with
higher energy fluency.
In conclusion, once again, neutrino radiance should be much larger than gamma one
in the most general hadronic jet model crossing star shells. However, the data show a comparable or a minor
neutrino radiance with respect to the gamma ones. This is the main need for a pure electronic jet in GRBs.

\section{Binary BH-NS feeding accretion disk and powering $\gamma$ jet}\label{sec:BH-NS}

In the light of this absence of GRB-$\nu$ (\ref{sec:lack}), we are forced to consider a new engine process able to avoid any pion decay chain. The most natural one is a binary system in empty space made by a neutron star (NS) and a black hole (BH) in an elliptical trajectory with each other.
At a nearby encounter, as depicted in Fig.~\ref{fig:01}, the NS may suddenly lose a fragment of its mass because of tidal forces at relativistic Roche limit \cite{1973ApJ...185...43F}. These neutrons are led within tens of minutes toward the last extended boundary ($r\sim3R_{\textrm{Schwarzschild}}$) of the BH while the decay $n\to p+e^-+\bar{\nu}_e$ takes place. The electrons will then escape at low MeV energy, leading to a poor spherical (hard to detect) signal, while the protons which don't gain too much energy, nor relevant momentum in the decay, will proceed in its geodetic spiraling in a disk-ring around the BH; the ring will be therefore a positive charged ring.
The almost relativistic electrons in the meantime will spread themselves in a nearly spherical fashion.
The neutron-proton coherent spiraling around the BH will then define a net positive charged current in a ring that is not compensated by a relativistic electronic component of the decay. This induces a huge axial magnetic field $\mathbf{B}_p$ proton-induced which is represented in Fig.~\ref{fig:02}; the magnetic lines force the electrons to concentrate themselves toward the BH accretion disk's poles (let's call them north and south according to the magnetic field polarity). The electrons will then be forced and squeezed by a powerful charged pump that accelerate the $e^-$ in a jet at highest energies well above the starting MeV ones. Within such a dense relativistic electron beam flow, because of self-electron Compton scattering, inverse Compton and pair production,  collinear pairs $e^+e^-$ and $\gamma$ will arise resulting in a final $\gamma$ jet. The BH spin and the ring spin will interact and precess among themselves.
  \begin{figure}[t]
 \centering
  \includegraphics[scale=.3]{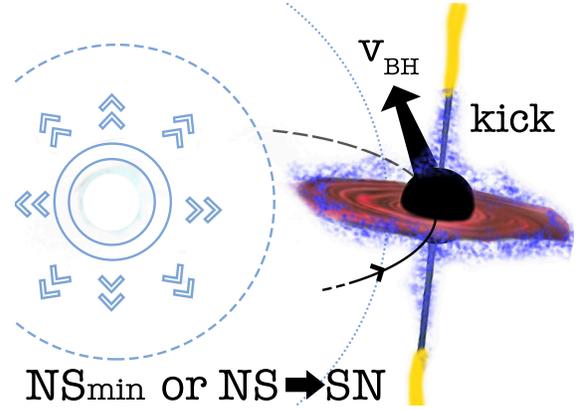}
\caption{Unstable NS explodes in a spherical SN-like event, observable days or weeks after first GRB blaze. Shells of energy of the supernova embrace the same BH jet. The asymmetric binary BH is suddenly without a companion and it is launched tangentially with a high speed kick (see \cite{2007ARep...51..308B}) in a fast flight holding alive its ring and its jet. The latest stages of the BH fed jet may shine as a SGR. The model NS-BH maybe dressed in a similar NS-NS evolution where the final relic is a spinning NS jet; this version may fit the SGRs or AX PSRs relics observed in our own galaxy.}
\end{figure}\label{fig:04}

In the proton disk, meanwhile, for the accumulated charged asymmetry, some of the external circuiting protons will start to escape at equatorial disk edges (see Fig.~\ref{fig:02}). Clearly, the extreme collimation of the pairs $e^+e^-$ and $\gamma$ avoids the Eddington opacity that normally occurs for spherical luminosity and the huge dense NS mass feeding the proton ring represents a very powerful engine ($\dot{m}_{\rm NS}\simeq10^{-6}\div10^{-5}\,\mathrm{M}_\odot\,\mathrm{s}^{-1}$).
This mass loss, then, powers the BH accretion disk and the jet, whose blazing toward the Earth is perceived as a GRB.
After a few days or months the NS is doomed; its strip for the benefit of the BH ring may lead to instabilities (see Fig.~\ref{fig:03}) and the reason for that is simple: a very minimal NS mass $(m^{\rm NS}_{min}\lesssim0.2\,\mathrm{M}_\odot)$ may become too light to hold together nuclei (see \cite{1998A&A...334..159S}) and its surface gravity weight becomes unable to compensate the nuclear chemical repulsion potential (as happens in a normal NS). Neutrons from the surface would then start to decay and escape making the degenerate system totally unstable in a matter of tens of seconds or few minutes (Fig.~\ref{fig:03}). This would lead to a sudden spherical  explosion appearing from Earth as a SN event (Fig.~\ref{fig:03b}). However, it is not trivial to tell if the critical minimal neutron star mass could release much more or much less energy than of a canonical SN. The energy potential budget for a NS collapsing in a normal SN accounts for around 10\% of the object rest mass ($\sim\mathrm{M}_\odot$). Therefore, an apparent SN-like event like the one celebrated SN-GRB related to GRB 980425 may be attributed to such  a simple process of a minimal NS explosion without any correlated beamed neutrino and with a few days (or a week, \cite{2007ApJ...658L...5M}) delay with respect to the main GRB blaze.
Naturally, the shining of the spherical NS explosion may heat and excite the external surrounding  (original SN shell from where NS itself or BH is formed) shell leading to spectroscopic emission and absorption lines that may mimic the SN explosion.
On the contrary, the Ni and/or the Co radioactive decay mode are not naturally born (therefore there might be a remarkable imprint to be discussed elsewhere that might distinguish the SN from the NS-like explosions).

We like to stress that this electromagnetic pump accelerator mechanism does not require any hadron parental engine, any consequent muons or  energetic neutrinos, explaining the the observed absence of ICECUBE  neutrino radiance (larger than the photon one) and-or the missing  GRB-$\nu$ correlation.

\subsection{Bimodal Short and long GRB}\label{sec:slGRB}

There are also natural corollary consequences of the proposed model: we can find a similar tale for a NS-SN binary collapse where one of the two NS ``eats'' and ``strips'' matter from the companion NS leading to a similar story-board.
 Because such a NS-NS binary systems are among the narrow ones then we imagine that also their characteristic blazing times are sharper
 leading to more short duration GRBs. Therefore these shorter GRBs may populate the short events, whose duration is below 2 s.

Larger sized BH-NS binaries, like the very recent candidate in the LIGO-VIRGO gravitational wave detection \cite{PhysRevLett.116.061102}, system may imply  a wider family of NS-BH with BH masses as large as $10\div100\div1000\,\mathrm{M}_\odot$. These are possibly the longer duration GRBs whose characteristic time is longer than 2 s.
The infrequent and sporadic presence  of largest BH makes rarer and rarer the longest GRB events explaining the rarest long life GRB , thousand second long.
  Also late  GRBs whose early explosion has not been in axis but whose late precessing jet is pointing (as a young SGRs) to us at a still high output, may appear as a short GRBs mostly at nearer cosmic distances (respect peaked GRB luminosity).

\section{Conclusions}\label{sec:end}

If the SGRB and LGRB  are explained by NS-NS (SGRB) and NS-BH (LGRB) models, then the main puzzle of the apparent over-Eddington luminosity is simply solved by high collimated beaming. The tidal ring-jet perturbation and the spinning of the BH versus the disk makes the jet spin and precessing as well as blaze in the observed almost  chaotic way (see Fig.~\ref{fig:04}).
The absence of longest events, almost comparable with largest optically violent variable quasar 3C 279 gamma flare is simply related with the rarity of super-massive BH (as the AGNs) respect lighter tens-of-hundreds or thousands solar masses. The coexistence of a SN-like event (for a quick review see i.e. \cite{2006ARA&A..44..507W, 2012arXiv1206.6979B}) is solved by light tidal NS sudden evaporation and consequent explosion. The absence of TeV neutrinos correlated with GRBs is guaranteed by the absence of any hadronic accelerator as well as leptonic neutrino tails in GRB. The thinner precessing jet moreover still explains the statistics we see, i.e. in Fig.~\ref{fig:06}.

The model consistence is based on the geometrical evolution of a thin persistent jet whose acceptance today, after twenty years, is becoming more and more obvious. We admit that for a long time we also assumed that such thin jets were powered by hadronic engine (muons) \cite{1999A&AS..138..507F, 2005NCimC..28..809F, 2009foap.conf..351F} and later on by their electron pairs fed by muons, \cite{2009astro2010S..75F, 2010MmSAI..81..440F, 2012MmSAI..83..312F}, but the absence of a $\nu$-$\gamma$ correlation and in particular the paucity of $\Phi_{\rm GRB}^\nu$ with respect to $\Phi_{\rm GRB}^\gamma$ forced us to the present ``neutron striptease'' jet-SN model, made by pure electron jets, mostly or totally free of hadronic engines.

 \acknowledgments
 We are in debt to Prof. Maxim Khlopov for deep  and useful comments, to Prof. B. Mele, Prof. P. Lipari, Prof. G. Salm\'e, Prof. R. Jatzen for their discussions and comments. This paper is dedicated to the memory of Ameglio Fargion, born on 2 June 1913, died on 22 April 2011.

\end{document}